\documentclass[a4paper]{article}

\usepackage{INTERSPEECH2022}
\usepackage{multirow}
\usepackage{booktabs,caption}
\usepackage[flushleft]{threeparttable}
\usepackage{color}
\usepackage[export]{adjustbox}
\usepackage{graphicx}
\usepackage{subcaption}
\usepackage{amssymb}
\title{FeaRLESS: Feature Refinement Loss for Ensembling Self-Supervised Learning Features in Robust End-to-end Speech Recognition}
\name{Szu-Jui Chen, Jiamin Xie, John H.L. Hansen}
\address{Center for Robust Speech Systems, University of Texas at Dallas, TX 75080} 
\email{ \{SzuJui.Chen, Jiamin.Xie, John.Hansen\}@utdallas.edu}

\begin{document}

\maketitle
\begin{abstract}
Self-supervised learning representations (SSLR) have resulted in robust features for downstream tasks in many fields. Recently, several SSLRs have shown promising results on automatic speech recognition (ASR) benchmark corpora. However, previous studies have only shown performance for solitary SSLRs as an input feature for ASR models. In this study, we propose to investigate the effectiveness of diverse SSLR combinations using various fusion methods within end-to-end (E2E) ASR models. In addition, we will show there are correlations between these extracted SSLRs. As such, we further propose a feature refinement loss for decorrelation to efficiently combine the set of input features. For evaluation, we show that the proposed ``FeaRLESS learning features" perform better than systems without the proposed feature refinement loss for both the WSJ and Fearless Steps Challenge (FSC) corpora.
\end{abstract}

%

\section{Introduction}
Deep neural networks (DNNs) have accelerated the progress in speech processing, thus enhancing the accessibility of speech-related technologies in daily life. Since DNNs are data hungry, researchers have been exploring ways to enlarge the model and use more transcribed data. However, collecting a sufficient amount of labeled data for training is generally not feasible. As a result, unsupervised and semi-supervised learning have been proposed for utilizing unlabeled data. Previous studies in computer vision (CV) and natural language processing (NLP) have shown promising results using unsupervised learning to learn representations from data \cite{gomez2017self, radford2018improving}. There is a similar trend in the speech field based on pre-trained models using unlabeled data for learning powerful speech features. Recently, a series of self-supervised representation learning models have been proposed. The extracted representations are usually referred to as self-supervised learning representation (SSLR), and have achieved state-of-the-art results on several benchmarks \cite{chang2021exploration, babu2021xls, hsu2021hubert, baevski2020wav2vec}.

In the study by Chang et al. \cite{chang2021exploration}, we see an advantage of using SSLRs over traditional hand-crafted features. With this observation, an obvious question arises: Would it be better to combine SSLRs for automatic speech recognition (ASR)? We explore an answer to this question by conducting experiments using alternate fusion methods for SSLRs. However, our preliminary results with the Wall Street Journal (WSJ) corpus showed that simple fusion methods do not improve word error rate (WER).

After further investigation, we suggest two possible reasons, where the combination of powerful SSLRs does not surpass the baseline. The first is that the WSJ corpus might be too simple for a fusion method to be beneficial. The second is that the extracted SSLRs have certain correlations with each other that lead to redundant information. In order to efficiently combine SSLRs, we propose a loss function to decorrelate features as a feature redundancy reduction method for end-to-end (E2E) ASR model. In addition, we further conduct the same experiments on the Fearless Steps Challenge (FSC) Phase 2 corpus which is spontaneous natural noisy speech.

The contribution of our work includes:
\begin{itemize}
    \item Investigate SSLR fusion in E2E ASR models
    \item Propose a feature refinement loss for efficient feature combining
\end{itemize}

\section{Background}
\subsection{Self-Supervised Learning Representations}
\label{SSLRs}
In this study, we consider objective function to categorize SSLRs, the main two being reconstructive loss and contrastive loss. For simplicity, we only note the SSLRs that are used in our experiments here, which are Hubert, Wav2Vec2.0, APC, TERA, and CPC. 

\noindent\textbf{Reconstructive} Models trained with reconstructive loss can either predict future frames conditioned on the past history, or they can predict current frames conditioned on the past and future information based on masking prediction concepts. SSLRs that belong to this category include autoregressive predictive coding (APC) \cite{chung2019unsupervised} and TERA \cite{liu2021tera}. 

\noindent\textbf{Contrastive} Alternatively, models trained with contrastive loss aim to distinguish positive from negative samples. SSLRs belonging to this category include contrastive predictive coding (CPC) \cite{oord2018representation} and Wav2Vec2.0 \cite{baevski2020wav2vec}.

Finally, an exceptional model named Hubert \cite{hsu2021hubert} was recently proposed that used masked language modeling and pseudo-labeling for speech representation learning. For more details of these SSLRs, please refer to \cite{chang2021exploration}.
\subsection{Feature Combination}
Earlier research studies have explored the effects of feature combination compared with model ensemble techniques in traditional hybrid systems \cite{schluter2007gammatone}. Various feature combination methods have been proposed, including linear discriminant analysis (LDA) \cite{zolnay2005acoustic} and concatenation, but only hand-crafted features such as MFCC and FBANK were explored. Recently, several studies have considered combining SSLRs with hand-crafted features \cite{chen2021scenario,vieting2021architectures}. To date however, none of these investigations have explored combinations of SSLRs. Hence, in this study we explore SSLR fusion experiments that utilize feature combination techniques and analyze their effectiveness for E2E speech recognition.

\begin{figure*}[t]
    \centering
    \includegraphics[width=0.7\linewidth]{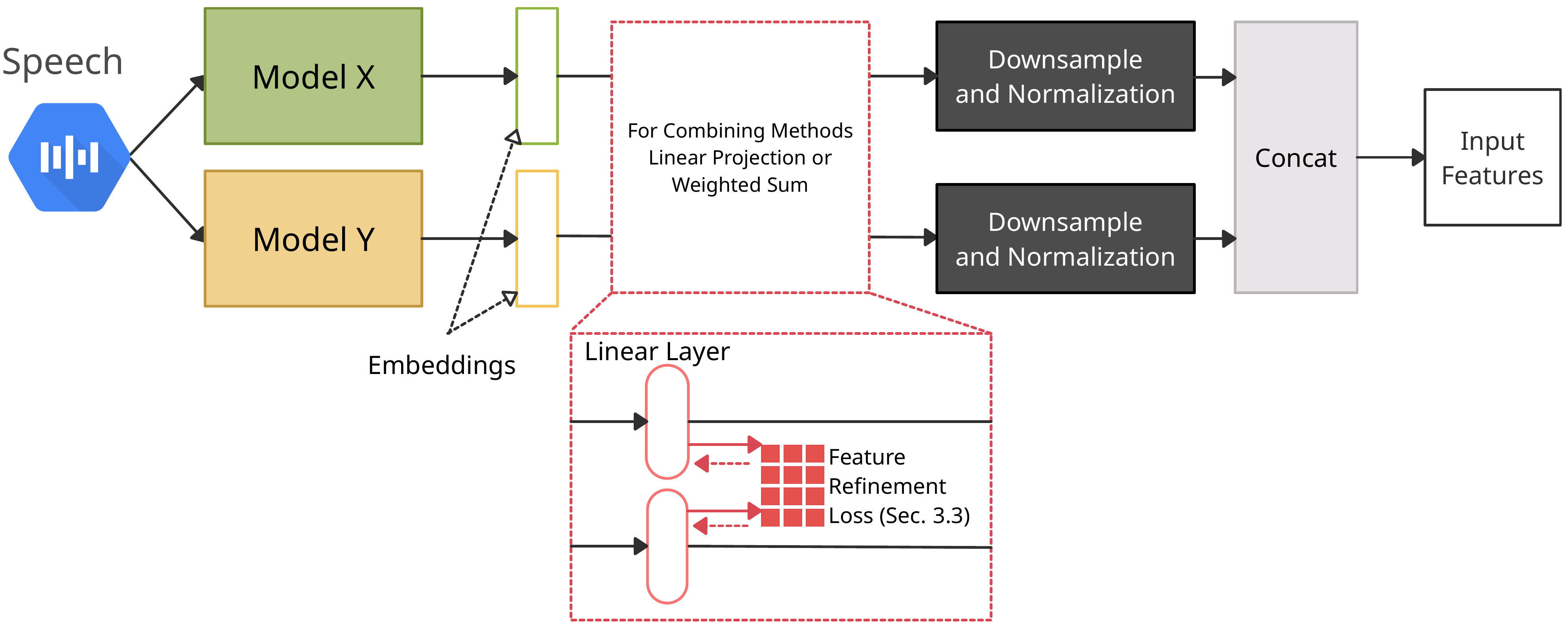}
    \caption{Frontend feature extractor flow chart}
    \label{fig:flow_chart}
\end{figure*}

\subsection{Decorrelation}
In this last area, the idea of decorrelation is noted, where earlier efforts have used this for model ensemble \cite{opitz2016efficient} and preventive overfitting \cite{cogswell2015reducing}. In the field of speech, past efforts explored feature decorrelation methods for speech recognition \cite{batlle1998feature, paliwal1999decorrelated, bao2013incoherent}, but only consider single feature decorrelation.
Inspired by these previous works, we explore advancements using decorrelation within a feature combination scenario for redundancy reduction. We propose a loss function for decorrelation between input features, in an effort to obtain a refined feature that is efficiently fused from SSLRs (details highlighted in Sec.~\ref{decorr}).

\section{Methods}
In this section, we introduce several methods to combine SSLR features. Without loss of generality, we describe the case of combining only two features, $\boldsymbol{U} \in {\rm I\! R^{T\times K_{1}}}$ and $\boldsymbol{V} \in {\rm I\! R^{T\times K_{2}}}$, where $T$ is the input feature length and $K_{1}$ and $K_{2}$ are the individual feature dimensions. The resolution of distinct features is kept the same via downsampling. Since SSLR features can consist of different dimension sizes, we introduce an additional affine transformation to unify the feature dimension size to a common lower dimension. This allows the combined feature to be flexibly mapped to any feature subspace of a downstream model. In the following sections, we will categorize combination methods by whether or not one requires an affine transformation.

\subsection{Combination Without Transformation}
\subsubsection{Concatenation}
\label{concat}
Concatenation is a natural approach for combining features without the need of unified dimension sizes. The original information in each SSLR feature is retained by concatenation. Consider an output feature $\boldsymbol{X} \in {\rm I\! R^{T\times (K_{1}+K_{2})}}$, our method can be written as:
\begin{align}
    \boldsymbol{X} &= 
        \begin{bmatrix}
        norm(\boldsymbol{U}), norm(\boldsymbol{V}) \\
        \end{bmatrix}, \label{concat}
\end{align}
where $norm(\cdot)$ is the mean normalization along time $T$, using the statistics from each utterance. The reason to normalize before combination methods is to remove any bias learned within individual SSLRs. Removing bias allows features to be on the same scale and benefits the subsequent combining mechanisms, fusing them to the subspace of a downstream task. 
It is worth noting that many previous studies \cite{sun2018face,shih2019real} use feature concatenation extensively.

\subsection{Combination With Transformation}
For a combination of features with varying dimensions, we introduce learnable parameters for transforming features to a common dimension size. Specifically, we let the network learn two affine transformations with specific weights and biases:
$\{\boldsymbol{W_{1}} \in {\rm I\! R^{K_{1}\times K}}$, $b_{1} \in {\rm I\! R^{1 x K}}$, $\boldsymbol{W_{2}} \in {\rm I\! R^{K_{2}\times K}}$, $b_{2} \in {\rm I\! R^{1 x K}}$\}. Using our previous notations, the transformed features are now $\tilde{\boldsymbol{U}} \in {\rm I\! R^{T\times K}}$ and $\tilde{\boldsymbol{V}} \in {\rm I\! R^{T\times K}}$, which are obtained as,
\begin{align}
    \tilde{\boldsymbol{U}} &= \boldsymbol{U}\cdot \boldsymbol{W_{1}} + b_{1}, \\
    \tilde{\boldsymbol{V}} &= \boldsymbol{V}\cdot \boldsymbol{W_{2}} + b_{2}.
\end{align}

\subsubsection{Linear Projection}
\label{LP}
Linear projection is the concatenation method with transformed features. Given an output feature $\boldsymbol{X} \in {\rm I\! R^{T\times 2K}}$, our method is written as,
\begin{align}
    \boldsymbol{X} &= 
        \begin{bmatrix}
        norm(\tilde{\boldsymbol{U}}), norm(\tilde{\boldsymbol{V}}) \\
        \end{bmatrix}, \label{lproj formula}
\end{align}
where $norm(\cdot)$ is the mean normalization along time $T$. Comparing with the output feature in Eq.~\ref{concat}, the feature transformation enables further control of the input features, such as for feature selection and refinement, which are described in Sec.~3.3.

\subsubsection{Weighted Sum}
\label{wsum}
In the weighted sum approach, transformed features are first scaled and then combined. The importance among input features is controlled by two learnable scalars $\alpha$ and $\beta$. For an output feature $X \in {\rm I\! R^{T\times K}}$, our method can be written as,

\begin{align}
    \boldsymbol{X} &=  \frac{1}{\alpha+\beta} \cdot \begin{bmatrix}
           norm(\tilde{\boldsymbol{U}}) & norm(\tilde{\boldsymbol{V}}) \\
            \end{bmatrix} \cdot \begin{bmatrix}
                                \alpha \\
                                \beta \\
                                \end{bmatrix},
\end{align}
where $norm(\cdot)$ is the mean normalization along time $T$. In comparison to linear projection in Eq.~\ref{lproj formula}, the weighted sum adds more parameters to control the combination, where prior knowledge can be applied if necessary.

\begin{figure*}[t]
  \centering
  \begin{subfigure}[]{0.35\textwidth}
    \centering
    \includegraphics[width=\textwidth]{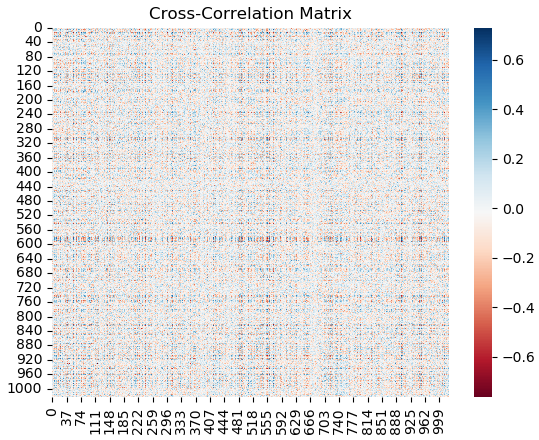}
    \caption{Before Linear Projection} 
    \label{fig2:a}
  \end{subfigure}
  \begin{subfigure}[]{0.32\linewidth}
    \centering
    \includegraphics[width=\linewidth]{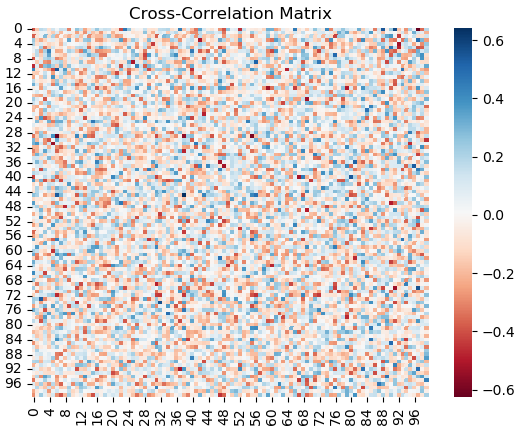} 
    \caption{After Linear Projection layer    \textbf{without} \\ feature refinement loss} 
    \label{fig2:b}
  \end{subfigure}
  \begin{subfigure}[]{0.32\linewidth}
    \centering
    \includegraphics[width=\linewidth]{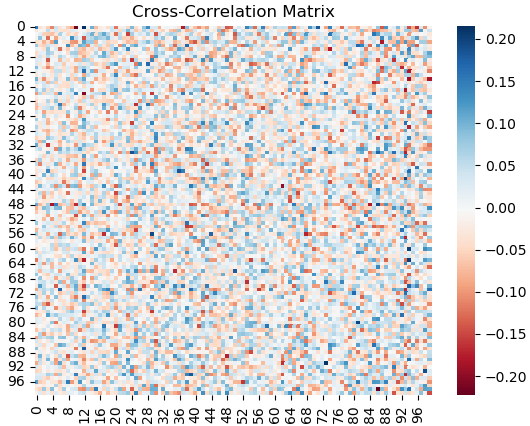} 
    \caption{After Linear Projection layer trained \\ \textbf{with} feature refinement loss} 
    \label{fig2:c}
  \end{subfigure}
  \vspace{-0.2cm}
  \caption{Cross-correlation matrix between Hubert and Wav2Vec2.0 of the utterance 4kac030m in WSJ dev03 set. The y-axis shows Hubert while the x-axis is Wav2Vec2.0. (a) uses raw features extracted from the pre-trained model. (b) and (c) compare the outcome of using a feature refinement loss, where (c) employs a threshold $\epsilon=0.2$.}
  \label{fig:corr matrix}
\end{figure*}

\subsection{Feature Refinement Loss}
\label{decorr}
The feature refinement loss guides the learning of feature transformations. 
As an ensemble strategy, the feature combination can benefit from a diverse set of input representations. However, we discover that non-trivial correlations can exist between SSLR features. For a random utterance, the correlation between extracted features from Hubert and Wav2Vec2.0 
can reach over 0.6, considered as a high correlation, as shown in Fig.~\ref{fig2:a}. It becomes even clearer in the lower dimensions after the linear projection layer, as shown in Fig.~\ref{fig2:b}. Since highly correlated features represent similar information, they are generally considered redundant for a system and especially for an ensemble. By introducing the feature refinement loss, we manage to reduce the correlation as low as 0.2, as shown in Fig.~\ref{fig2:c}. We found that it is an essential procedure to diminish redundancy between SSLR features for combination.

Our proposed feature refinement loss is calculated from the cross-correlation matrix $\boldsymbol{C} \in {\rm I\! R^{K\times K}}$ between two transformed feature matrices. Thus, the optimization of the loss is equivalent to reducing correlations between input features. Using our previous notations, the cross-correlation matrix is obtained by,
\begin{equation}
    \boldsymbol{C} = norm(\tilde{\boldsymbol{U}})^{T} \cdot norm(\tilde{\boldsymbol{V}}), \\
\end{equation}
where $norm(\cdot)$ is the mean and \textit{variance} normalization along time $T$. Note that the pre-normalization ensures the correlation is bounded between -1 and 1.

The feature refinement loss is then defined by, 
\begin{equation}
    L_{refine} \overset{\Delta}{=}
    \sum_{i=1}^{K}{\sum_{j=1}^{K}{
    \begin{cases}
        (C_{ij})^{2} & if\ |C_{ij}| > \epsilon \\
        0 & otherwise
    \end{cases}}}, \\
\end{equation}
where $\epsilon$ is a threshold parameter. For example, when $\epsilon = 0.2$, any input entries that results in a correlation between -0.2 and 0.2 is masked from the loss. For batch mode training, the loss is averaged along the batch dimension.

The final loss is a combination of ASR loss and feature refinement loss with a hyperparameter $\lambda$ for scaling the latter:
\begin{equation}
        L = L_{asr} + \lambda \cdot L_{refine}. \\
\end{equation}
Note that the feature refinement loss will only affect the linear projection layer during the E2E ASR model training as shown in Fig.~\ref{fig:flow_chart}, whereas the ASR loss impacts the entire architecture except the frozen pre-trained SSLR models.

\section{Experiments}
To evaluate the effectiveness of SSLR combinations and feature refinement loss, we first perform experiments on the Wall Street Journal (WSJ) corpus, followed by testing on Fearless Steps Challenge (FSC) phase 2 corpus. The ESPnet toolkit \cite{watanabe2018espnet} is used for all experiments. The SSLRs used in our experiments were summarized in Sec.~\ref{SSLRs}.

\subsection{WSJ \& Fearless Steps Challenge Corpora}
The WSJ corpus \cite{paul1992design} consists of 81 hours of read newspaper speech. The standard subset si284, dev93, and eval92 are used for training, development, and testing in out experiments.

The FSC phase 2 corpus \cite{joglekar2020fearless,hansen2018fearless} is a subset of the 19,000-hour original Fearless Steps Corpus. FSC-2 is the original Apollo conversational speech with labeled meta-data that consists of 5 selected channels: Network Controller (NTWK), Electrical, Environmental and Consumables Manager (EECOM), Guidance Navigation and Control (GNC), Flight Director (FD), and Mission Operations Control Room (MOCR). There are distinct acoustic characteristics(e.g., noise, distortion, background inference, etc.) within these channels. In this work, we conduct experiments on ASR track 2 with segmented utterance level transcriptions. The training, development, and evaluation sets are 28 hours, 7.6 hours, and 10.6 hours respectively.

\subsection{Setups}
We use SSLR models with frozen parameters to extract features on the fly. Extracted features are first projected to 100 dimensions if using linear projection (Sec.~\ref{LP}) or weighted sum (Sec.~\ref{wsum}). Next, features are downsampled if we are combining features with different time strides --- note that the time stride for Hubert and Wav2Vec2.0 is 20ms, while all other SSLRs have a stride of 10ms. After that, features are normalized across time using statistics from each utterance and then concatenated or summed along feature dimension. The combined features are further transformed to the 80-dimensional subspace of ASR task by a linear layer. Finally, these processed features are fed into the E2E ASR model, consisting of a 12-layers Conformer \cite{gulati2020conformer} encoder followed by a 6-layers Transformer decoder.

When using the proposed feature refinement loss, the loss weight is set to 0.3 with the threshold parameter $\epsilon=0.2$ for WSJ, and a loss weight 0.005 with $\epsilon=0.6$ for FSC corpus. We use a warm-up learning scheduler. It has a peak learning rate 0.002 and warm-up steps 10,000 for WSJ while a peak learning rate 0.002 and warm-up steps 18,000 for FSC corpus. WSJ experiments use a language model (LM), while the FSC experiments do not.


\subsection{Results and Analysis}
\subsubsection{Results of Combining SSLR}
In Table~\ref{tab:t1 combinations}, we show word error rate (WER) performance for combinations of different SSLRs using the concatenation method on WSJ. We found that individually powerful SSLRs, such as Hubert and Wav2Vec2.0, drive the combination results. This can be seen when combining TERA with Wav2Vec2.0 instead of APC and combining CPC with Wav2Vec2.0 instead of APC. If Wav2Vec2.0 is included, the result is better. However, TERA or CPC combined with Wav2Vec2.0 only outperforms Wav2Vec2.0 in the eval92 set. When combining SSLRs of comparable strength, better results typically occur. For example, combining TERA or CPC with APC is better than using APC alone, or combining Hubert with Wav2Vec2.0 is also better than individually. In all, we obtain the best WER when combining Hubert with Wav2Vec2.0 SSLRs.

However, we also observe that adding a LM might actually reduce improvements using combination. For example, the Hubert and Wav2Vec2.0 combination improves only the dev93 set after introducing LM. Although the LM does reduce the benefit of SSLR combinations, we can still conclude that combining SSLRs can be beneficial for WER performance.

\begin{table}[t]
  \caption{WER of different combinations of SSLRs using concatenation on WSJ corpus.}
  \label{tab:t1 combinations}
  \begin{tabular}{ l c c | c c}
    \toprule
    \multirow{2}{*}{\centering Model} & \multicolumn{2}{c}{\textbf{With LM (\%)}} & \multicolumn{2}{c}{\textbf{No LM (\%)}} \\ 
    & dev93 & eval92 & dev93 & eval92 \\
   	\midrule
    Hubert &  3.0 & \textbf{1.6} & 4.6 & 3.5 \\
    Wav2Vec2.0 & 3.0 & 2.1 & 4.8 & 3.7\\
    TERA & 6.4 & 4.0 & 10.8 & 8.8 \\
    CPC & 6.6 & 3.9 & 11.5 & 8.9 \\
    APC & 7.2 & 4.4 & 12.0 & 8.7 \\
    \midrule
    Hubert + Wav2Vec2.0 & \textbf{2.8} & \textbf{1.6} & \textbf{4.4} & \textbf{3.4}\\
    TERA + Wav2Vec2.0 & 3.2 & 1.8 & 4.8 & 3.6\\
    CPC + Wav2Vec2.0 & 3.2 & 2.0 & 5.2 & 4.2 \\
    TERA + APC & 6.8 & 3.8 & 11.8 & 8.4 \\
    CPC + APC & 6.5 & 4.5 & 11.6 & 8.9 \\    
    \bottomrule
  \end{tabular}
\end{table}

\subsubsection{Combination Methods Comparison}
Next, in Table~\ref{tab:t2 wsj results} we compare WER performances on WSJ under different fusion methods with the Hubert and Wav2Vec2.0 combination. Both concatenation and weighted sum achieve similar performance scores, with only a slight difference on the dev93 set when a LM is not applied. However, linear projection performs worse on both dev93 and eval92 sets versus the other two methods with a LM included. Finally, adding feature refinement loss in addition to linear projection achieves 2.8\% and 1.5\% WER on dev93 and eval92 sets, which is a +6.67\% and +6.25\% relative improvement compared to Hubert only. Note that simply applying the feature refinement loss on top of the linear projection method provides a 6.67\% and 11.76\% relative improvement on WER. We also discover that the $\alpha$ in weighted sum method is 0.68 for Hubert and the $\beta$ is 0.32 for Wav2Vec2.0, which matched the WER performance of these two SSLRs in Table~\ref{tab:t1 combinations}, showing Hubert is a more important SSLR feature.

\begin{table}[h]
\scalebox{0.85}{
\begin{threeparttable}
  \caption{WER on WSJ corpus with fusion methods applied to Hubert and Wav2Vec2.0 combination. H, W, and LP stand for Hubert, Wav2Vec2.0, and linear projection respectively.}
  \label{tab:t2 wsj results}
  \begin{tabular}{ l l c c | c c }
    \toprule
    \multirow{2}{*}{\centering Model} & \multirow{2}{*}{\centering Fusion Method} & \multicolumn{2}{c}{\textbf{With LM (\%)}} & \multicolumn{2}{c}{\textbf{No LM (\%)}}\\
    & & dev93 & eval92 & dev93 & eval92 \\
   	\midrule
   	H & - & 3.0 & 1.6 & 4.6 & 3.5 \\
   	H + W & Concatenation & \textbf{2.8} & 1.6 & 4.4 & 3.4\\
   	H + W & Weighted Sum & \textbf{2.8} & 1.6 & \textbf{4.3} & 3.4 \\
    H + W & LP & 3.0 & 1.7 & - & - \\
    H + W & LP + Refinement loss & \textbf{2.8} & \textbf{1.5} & 4.4 & \textbf{3.3} \\
    \bottomrule
  \end{tabular}
\end{threeparttable}}
\end{table}

In Table~\ref{tab:t3 fsc results}, we extend our experiment to the FSC corpus. Using linear projection to combine Hubert and Wav2Vec2.0 gives +5.57\% and +7.88\% relative WER improvements on the eval set comparing Hubert and Wav2Vec2.0 respectively. After applying feature refinement loss, we further see an -0.32\% and +0.59\% relative WER improvement for dev and eval sets. These results show that our proposed feature refinement loss can be generalized to a noisy naturalistic real-world corpus.

\begin{table}[t]
  \caption{WER of FSC corpus with fusion methods applied to Hubert and Wav2Vec2.0 combination. H, W, and LP stand for Hubert, Wav2Vec2.0, and linear projection respectively. Results here are without language model.}
  \label{tab:t3 fsc results}
  \begin{tabular}{ l l c c}
    \toprule
    \multirow{2}{*}{\centering Model} & \multirow{2}{*}{\centering Fusion Method} & \multicolumn{2}{c}{\textbf{FSC (\%)}} \\
    & & dev & eval \\
   	\midrule
    Hubert & - & 33.2 & 35.9 \\
    Wav2Vec2.0 & - & 33.8 & 36.8 \\
    Hubert + Wav2Vec2.0 & LP & \textbf{31.2} & 33.9 \\
    Hubert + Wav2Vec2.0 & LP + Refinement loss & 31.3 & \textbf{33.7} \\
    \bottomrule
  \end{tabular}
\end{table}

\section{Conclusion}
In this paper, we have explored several combinations of SSLRs using linear projection, concatenation, and weighted sum methods in an end-to-end ASR framework. With our observation reflecting correlations between features, we further proposed a feature refinement loss for redundancy reduction when combining features. Experiments conducted on WSJ and FSC corpora show clear improvements when combining SSLRs and using the feature refinement loss. We also observed a +6.25\% and +6.13\% relative improvement compared to Hubert only for the WSJ eval92 and FSC eval sets. Our future work will include exploring other powerful SSLRs, combining additional features, and experimenting with alternative feature refinement approaches.

\section{Acknowledgement}
This project was funded, in part, by NSF-CISE Award 2016725, and partially by the University of Texas at Dallas from the Distinguished University Chair in Telecommunications Engineering held by J. Hansen. The authors would like to express our sincere thanks for valuable discussion with Huang-Cheng Chou.

\bibliographystyle{IEEEtran}

\bibliography{mybib}

\end{document}